\documentclass[twocolumn,prb]{revtex4}
\usepackage{amsmath,amsbsy,amssymb,graphicx}
\usepackage{times}

\begin{document}

\title{ Hourglass Fermion Surface States in Stacked Topological Insulators \\
with Nonsymmorphic symmetry}
\author{Motohiko Ezawa}
\affiliation{Department of Applied Physics, University of Tokyo, Hongo 7-3-1, 113-8656,
Japan}

\begin{abstract}
Recently a nonsymmorphic topological insulator was predicted, where the
characteristic feature is the emergence of a "hourglass fermion" surface
state protected by the nonsymmorphic symmetry. Such a state has already been
observed experimentally. We propose a simple model possessing the hourglass
fermion surface state. The model is constructing by stacking the
quantum-spin-Hall insulators with the interlayer coupling introduced so as
to preserve the nonsymmorphic symmetry and the time reversal symmetry. The
Dirac theory is also derived, whose analytical results reproduce the
hourglass fermion surface state remarkably well. Furthermore, we discuss how
the hourglass state is destroyed by introducing perturbations based on the
symmetry analysis. Our results show that the hourglass fermion surface state
is universal in the helical edge system with the nonsymmorphic symmetry.
\end{abstract}

\maketitle

\begin{figure}[t]
\centerline{\includegraphics[width=0.47\textwidth]{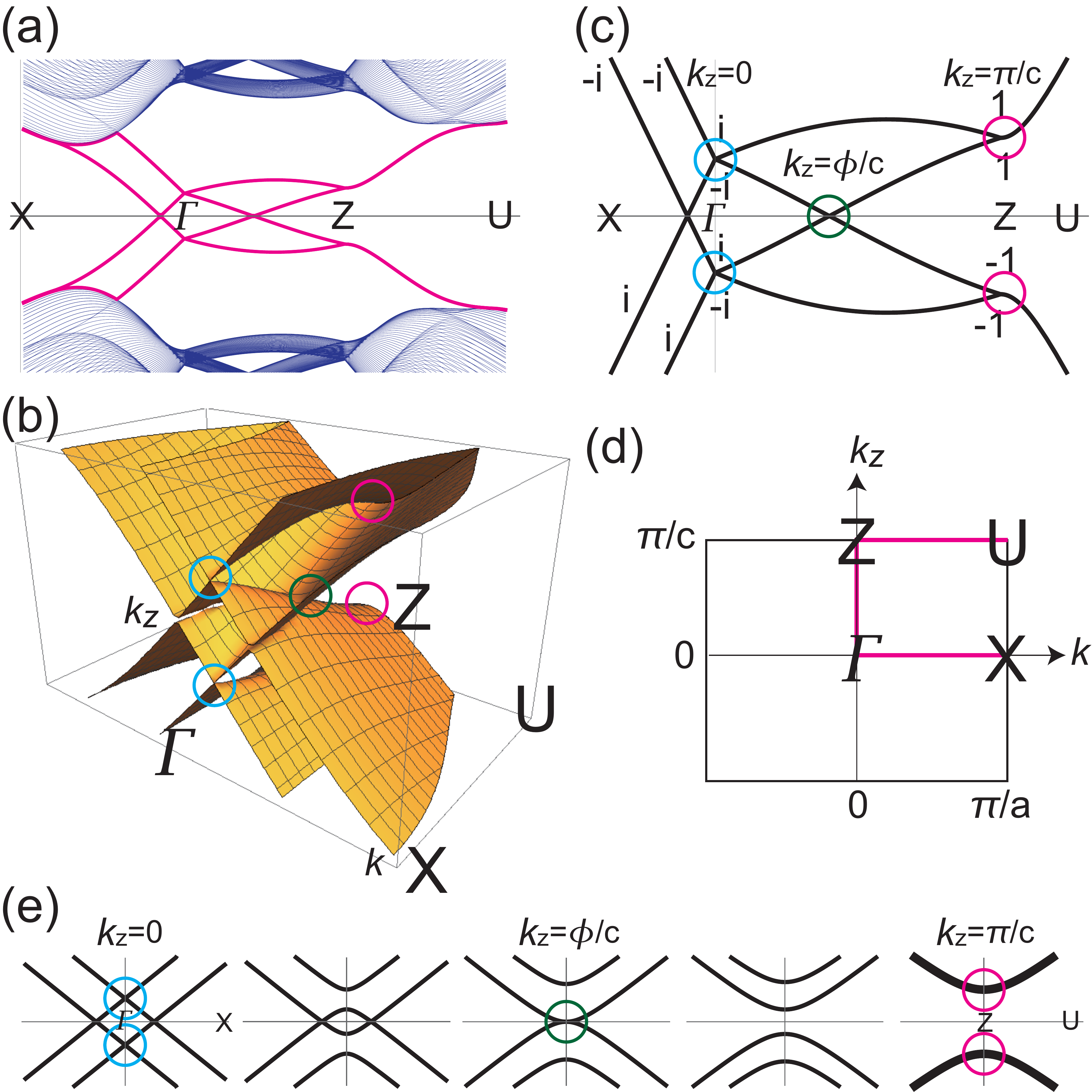}}
\caption{Band structure of the glided QSH insulator. (a) The band structure
along the X$\Gamma $ZU line. The red curves near the Fermi level represent
the surface state. (b) Bird's eye's views of the energy spectrum for the
surface state in the $(k,k_{z})$ plane. There are two vertically placed
Dirac cones at the $\Gamma $ point, representing the hourglass fermion
surface state. The nonsymmorphic symmetry protected gap closing point is
marked by a green circle. The Kramers (pseudo Kramers) doublets are marked
by cyan (magenta) circles. (c) The energy spectrum based on the Dirac theory
corresponding to (a), reproducing well the results for the surface state.
The eigenvalues of the glide operator are shown. (c) The Brillouin zone for
the surface, where X= $(\protect\pi /a,0)$, $\Gamma =(0,0)$, Z=$(0,\protect%
\pi /c)$, and U=$(\protect\pi /a,\protect\pi /c)$ in the $(k,k_{z})$ plane.
Here, $a$ and $c$ are the lattice constants of the honeycomb lattice and the 
$k_{z}$ direction, respectively. (e) The band structure along the $k$
direction for various $k_{z}$. }
\label{FigMain}
\end{figure}

\section{Introduction}

Topological insulators are fascinating concept found in condensed matter
physics\cite{Hasan,Qi}. They can be either genuine such as quantum-Hall and
quantum-anomalous-Hall (QAH) insulators\cite{Haldane,Chang} or symmetry
protected. Time-reversal symmetry (TRS) protected topological insulators
such as quantum-spin-Hall (QSH) insulators evoke intense research on this
topic\cite{KM,B1,B2}. Later they are generalized to topological crystalline
insulators, where the topology is protected by the crystalline symmetry such
as the mirror symmetry\cite{TCI,Hsieh,Tanaka,Dziawa,Xu,Robert,Ando}. However
the symmetry is restricted to be symmorphic although some extension to
combined symmetry is explored\cite{Aris}.

Recently, topological crystalline insulators are generalized further to
include the nonsymmorphic symmetry, and called nonsymmorphic topological
crystalline insulators\cite%
{Van,Para,Fang,Dong,Young,Watanabe,Po,BJYang,Murakami,Sch,ShiozakiB,Shiozaki,Chen,Varjas,Po}
. A nonsymmorphic topological insulator was predicted in KHgSb by
first-principles calculations\cite{Hour,Coho} and experimentally observed by
ARPES (Angle-Resolved PhotoEmisison Spectroscopy)\cite{KHg} in the same
material soon after. A prominent feature is that there emerges entirely a
new surface state called "hourglass fermion". It consists of four bands, in
which two bands cross along a high-symmetry line. This band crossing is
protected by the nonsymmorphic symmetry. The constructed effective theory is
comprised of complicated forms based on three orbitals of Sb and one orbital
of Hg, and is applicable only in the vicinity of the certain high-symmetry
points\cite{Hour,Coho}.

In this paper, motivated by these works\cite{Hour,Coho,KHg}, we propose a
tight-binding model producing a hourglass fermion surface state [Fig.\ref%
{FigMain}] by the stacked helical edge states with the nonsymmorphic
symmetry. They are realized by the binary stacking of the QSH insulators.
The unit cell contains two layers of the QSH insulators, which results in a Z%
$_{2}$-trivial insulator with respect to the TRS. Each layer has helical
edge modes; two right-going edge modes with up spin and two left-going edge
modes with down spin. In general, these helical edge modes are gapped out by
interlayer couplings. Nevertheless, an interesting feature is that a gap
closing is assured even after introducing the interlayer couplings provided
the nonsymmorphic symmetry is intact, leading to a hourglass fermion surface
state in the stacked system. (Let us call it the hourglass state for
simplicity.) To generate such a state, we first make a half lattice
transformation $\overline{t}_{z}=e^{-ick_{z}/2}$, where $c$ is a lattice
constant along the $\mathbf{z}$ direction\cite{Young,Shiozaki,Fang,Coho}.
Then we perform an additional mirror reflection. The combined operation is
the glide operation. We also derive the glided Dirac theory as an effective
theory for the hourglass state. It accounts for the above mentioned
properties of the hourglass state analytically [Fig.\ref{FigMain}(c)]. We
also investigate the breaking of the hourglass state by introducing
perturbations possessing various symmetries. Our results show that the
hourglass state is universal in the helical edge system with the
nonsymmorphic symmetry.


\section{Glided QAH insulator}

We start with an explicit construction of a spinless tight-binding model
possessing a nonsymmorphic symmetry protected surface state. We propose to
stack the QAH insulators so as to preserve the nonsymmorphic symmetry. The
QAH insulator is described by the Haldane Hamiltonian\cite{Haldane}, 
\begin{equation}
\hat{H}_{\text{QAH}}=-t\sum_{\left\langle i,j\right\rangle }c_{i}^{\dagger
}c_{j}+i\frac{\lambda _{\text{H}}}{3\sqrt{3}}\sum_{\left\langle
\!\left\langle i,j\right\rangle \!\right\rangle }\nu _{ij}c_{i}^{\dagger
}c_{j},
\end{equation}%
where $\left\langle \!\left\langle i,j\right\rangle \!\right\rangle $ run
over all the next-nearest neighbor hopping sites and $\nu _{ij}=\left( \vec{d%
}_{i}\times \vec{d}_{j}\right) /\left\vert \vec{d}_{i}\times \vec{d}%
_{j}\right\vert $ with $\vec{d}_{i}$ and $\vec{d}_{j}$ the two nearest bonds
connecting the next-nearest neighbors. The first term represents the usual
nearest-neighbor hopping on the honeycomb lattice with the transfer energy $%
t $ and the second term represents the Haldane interaction with the strength 
$\lambda _{\text{H}}$. We propose the glided QAH insulator model, which is $%
\hat{H}_{\text{gQAH}}=\sum_{\mathbf{k}}H_{\text{gQAH}}\left( \mathbf{k}%
\right) c_{\mathbf{k}}^{\dagger }c_{\mathbf{k}}$ in the momentum space with 
\begin{align}
H_{\text{gQAH}}\left( \mathbf{k}\right) & =H_{\text{QAH}}\left(
k_{x},k_{y}\right) +G_{\alpha }\left( k_{z}\right) H_{\text{H}}\left(
k_{x},k_{y}\right) G_{\alpha }\left( k_{z}\right) ^{-1}  \notag \\
& +\Gamma _{\text{QAH}}\left( k_{z}\right) ,
\end{align}%
where $\alpha =x$ or $y$. Hereafter we choose $\alpha =x$ for definiteness.
The first term is the Haldane Hamiltonian for the $A$ layer. The second term
is the Hamiltonian for the $B$ layer constructed with the aid of the glide
operator $G_{x}\left( k_{z}\right) $. The third term $\Gamma _{\text{QAH}%
}\left( k_{z}\right) $ represents the interlayer coupling preserving the
nonsymmorphic symmetry.

The glide operation $G_{x}\left( k_{z}\right) $\ is the successive operation
of the half translation $\overline{t}_{z}$ and the mirror operation $M_{x}$.
The mirror operation $M_{x}$ is just the reflection $\mathcal{R}_{x}$ in the
absence of the spin, $\mathcal{R}_{x}:\left( k_{x},k_{y},k_{z}\right)
\rightarrow \left( -k_{x},k_{y},k_{z}\right) $. It is given by\cite%
{Fang,Sch,Shiozaki} 
\begin{equation}
G_{x}\left( k_{z}\right) =e^{-i\frac{ck_{z}}{2}}\Omega (k_{z})\mathcal{R}_{x}
\label{glide0}
\end{equation}%
with%
\begin{equation}
\Omega (k_{z})=\cos \frac{ck_{z}}{2}\eta _{x}+\sin \frac{ck_{z}}{2}\eta _{y}.
\label{Omega}
\end{equation}%
Here, $\eta _{\alpha }$ are the Pauli matrices for the layer degrees of
freedom (pseudospin). It follows that $G_{x}^{2}\left( k_{z}\right)
=e^{-ick_{z}}\eta _{0}$.

\begin{figure*}[t]
\centerline{\includegraphics[width=0.75\textwidth]{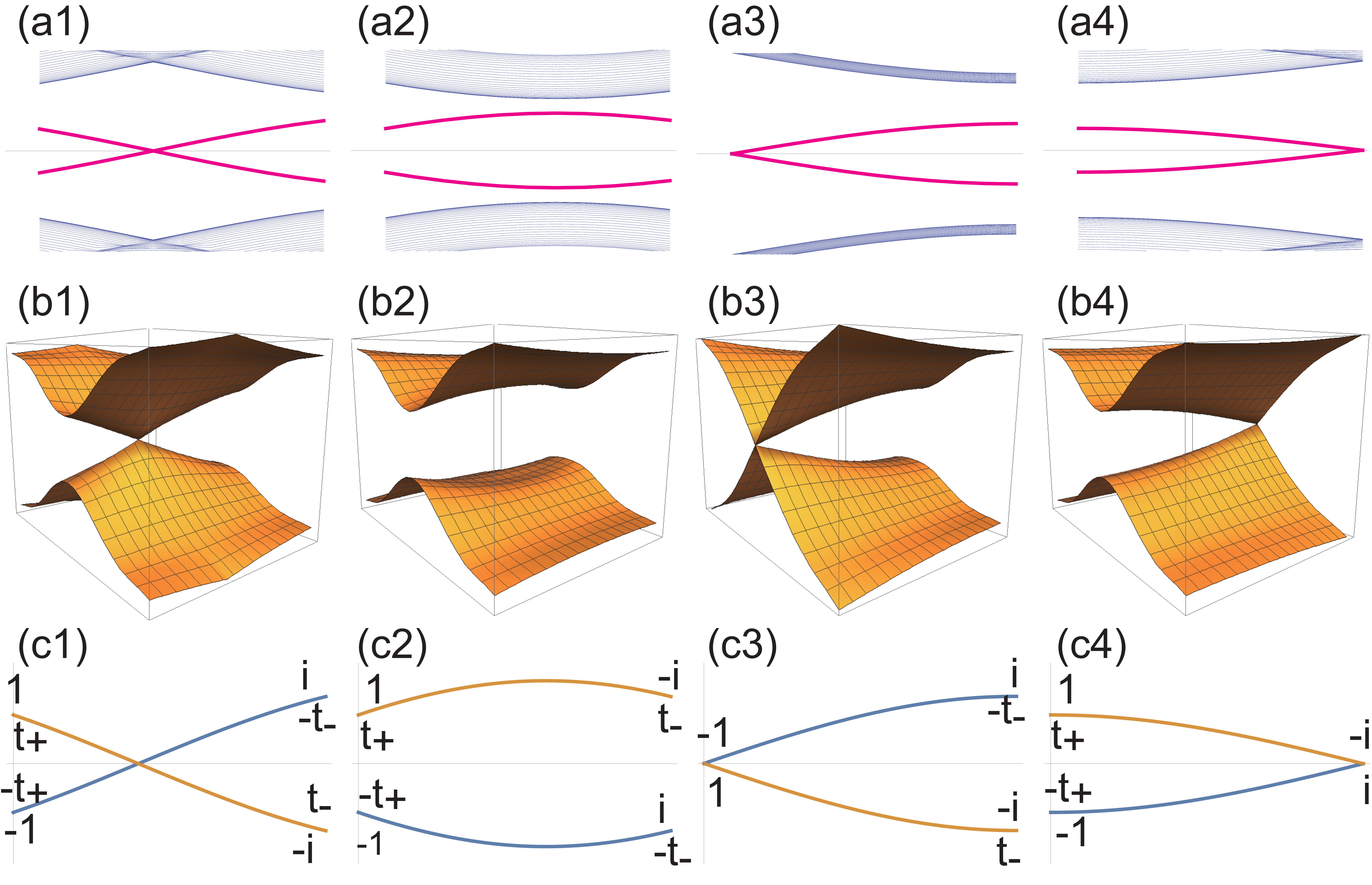}}
\caption{Band structure of the glided QAH insulator model. The band
structure along the line $k=0$ for $0\leq ck_{z}\leq \protect\pi $: (a1) $%
t_{+}t_{-}<0$, (a2) $t_{+}t_{-}>0$, (a3) $t_{+}=0$, and (a4) $t_{-}=0$. The
red curves near the Fermi level represent the surface states. (b1)$\sim $
(b4) Bird's eye's views of the energy spectrum for the surface states
corresponding to (a1)$\sim $(a4) in the $(k,k_{z})$ plane. (c1)$\sim $(c4)
The energy spectrum based on the Dirac theory corresponding to (a1)$\sim $
(a4). The eigenvalues of the glide operator are shown. }
\label{FigFour}
\end{figure*}

We determine the term $\Gamma _{\text{QAH}}\left( k_{z}\right) $. The only
condition imposed on it is $G_{x}\left( k_{z}\right) \Gamma _{\text{QAH}%
}\left( k_{z}\right) G_{x}^{-1}\left( k_{z}\right) =\Gamma _{\text{QAH}%
}\left( k_{z}\right) $, or 
\begin{equation}
\Omega (k_{z})\Gamma _{\text{QAH}}\left( k_{z}\right) \Omega
^{-1}(k_{z})=\Gamma _{\text{QAH}}\left( k_{z}\right) .
\end{equation}%
The simplest solution is obviously given by 
\begin{equation}
\Gamma _{\text{QAH}}\left( k_{z}\right) =f\left( k_{z}\right) \Omega (k_{z})
\end{equation}
with an arbitrary c-number function $f\left( k_{z}\right) $. We choose $%
f\left( k_{z}\right) $ to represent the coupling between the adjacent
layers, or%
\begin{equation}
\Gamma _{\text{QAH}}\left( k_{z}\right) =\left( t_{+}\cos \frac{ck_{z}}{2}%
+t_{-}\sin \frac{ck_{z}}{2}\right) \Omega (k_{z}),  \label{GammaH}
\end{equation}%
where $t_{+}$ ($t_{-}$) represents an ordinary (skew) interlayer hopping
amplitude.

The glide operation $G_{x}\left( k_{z}\right) $\ acts on the Hamiltonian as 
\begin{equation}
G_{x}\left( k_{z}\right) H_{\text{gQAH}}\left( k_{x},k_{y},k_{z}\right)
G_{x}^{-1}\left( k_{z}\right) =H_{\text{gQAH}}\left(
-k_{x},k_{y},k_{z}\right) .  \label{EqA}
\end{equation}%
We focus on the surface made of the edges of each layers, where $k_{y}=0$
and we set $k=k_{x}$.

It follows from Eq.(\ref{EqA}) that the glide operation commutes with the
Hamiltonian for the glide invariant plane $k=0$, 
\begin{equation}
\left[ H_{\text{gQAH}}\left( 0,k_{z}\right) ,G_{x}\left( k_{z}\right) \right]
=0.
\end{equation}
The operation of $G_{x}\left( k_{z}\right) $ twice results in the one-unit
cell translation. The eigen function $\left\vert \psi ^{\pm }\left(
0,k_{z}\right) \right\rangle $ satisfies 
\begin{equation}
G_{x}\left\vert \psi ^{\pm }\left( 0,k_{z}\right) \right\rangle =g_{\pm
}\left( k_{z}\right) \left\vert \psi ^{\pm }\left( 0,k_{z}\right)
\right\rangle .
\end{equation}
The eigenvalues are 
\begin{equation}
g_{\pm }\left( k_{z}\right) =\pm e^{-ick_{z}/2}
\end{equation}
since $G_{x}^{2}\left( k_{z}\right) =e^{-ick_{z}}\eta _{0}$. Especially $%
g_{\pm }\left( 0\right) =\pm 1$, $g_{\pm }\left( \pi /c\right) =\pm i$. All
bands on the high-symmetry line $k=0$ is labeled by the eigenvalues of $%
G_{x} $. We show the glide eigenvalues $g_{\pm }\left( k_{z}\right) $\ in
Figs.\ref{FigFour} (c1)$\thicksim $(c4).

We plot the band structure along the $k_{z}$ axis in Figs.\ref{FigFour}(a1)$%
\thicksim $(A4) for typical values of $t_{\pm }$. The isolated curves marked
in red represent the surface modes. They are present between the bulk gap.
We show the bird's eye's views of the band structure of the surface modes in
the $(k,k_{z})$ plane in Figs.\ref{FigFour}(b1)$\thicksim $(b4). There are
four types of the surface modes depending on $t_{+}$ and $t_{-}$. They are
classified as follows:\newline
1) When $t_{+}t_{-}<0$ they touch each other at the Fermi level at a certain
point $(0,k_{z}^{0})$. \newline
2) When $t_{+}t_{-}>0$, they never touch. \newline
3) When $t_{+}=0$, they touch at $(0,0)$. \newline
4) When $t_{-}=0$, they touch at $(0,\pi /c)$. \newline
This classification is made clear based on the effective theory valid near
the Fermi level, as we see soon.

\section{Glided Dirac theory of chiral edges}

We construct the Dirac theory in order to obtain deeper understanding of the
surface state. The chiral edge state of the glided QAH insulator is given by 
$H_{0}=vk\eta _{z}$ with $v\propto \lambda _{\text{H}}$ and $G_{x}\eta
_{z}G_{x}^{-1}=-\eta _{z}$. On the other hand it is impossible to take the
continuum limit in the $k_{z}$ direction. Hence the Hamiltonian for the
stacked chiral edges with the nonsymmorphic symmetry is given by%
\begin{equation}
H_{\text{gQAH}}=vk\eta _{z}+\Gamma _{\text{QAH}}\left( k_{z}\right) ,
\end{equation}%
together with Eq.(\ref{GammaH}). The energy spectrum reads 
\begin{equation}
E_{\pm }\left( k,k_{z}\right) =\pm \sqrt{v^{2}k^{2}+\left(
t_{+}^{2}+t_{-}^{2}\right) \sin ^{2}\frac{ck_{z}-\phi }{2}},  \label{EneSpeH}
\end{equation}%
where $\tan \frac{\phi }{2}=-t_{-}/t_{+}$. Especially we find 
\begin{equation}
E_{\pm }\left( 0,0\right) =\pm t_{+},\qquad E_{\pm }\left( 0,\pi /c\right)
=\pm t_{-}.
\end{equation}%
We plot the energy spectrum in Figs.\ref{FigFour}(c1)$\thicksim $ (c4),
which agrees with the result in Figs.\ref{FigFour}(a1)$\thicksim $(a4)
determined by the tight-binding model remarkable well.

The above mentioned classification of the surface modes is simply derived
from the analytic formula Eq.(\ref{EneSpeH}). In particular, by solving $%
E_{\pm }\left( k,k_{z}\right) =0$, we find $\left( k,k_{z}\right) =(0,\phi
/c)$, which is the gap closing point in Fig.\ref{FigFour}(b1). First, the
appearance of the gapless modes is protected by the nonsymmorphic symmetry.
Second, they appear at the same point in the $\left( k,k_{z}\right) $ plane
due to the chiral symmetry,%
\begin{equation}
CH_{\text{gQAH}}(k,k_{z})C^{-1}=-H_{\text{gQAH}}(-k,k_{z}),
\label{ChiralSymm}
\end{equation}%
where $C=\eta _{z}$ is the chiral operator. We can check $C\Omega
(k_{z})C^{-1}=-\Omega (k_{z})$ for Eq.(\ref{Omega}), from which the relation
Eq.(\ref{ChiralSymm}) follows.

\section{Glided QSH insulator}

We proceed to introduce the spin degrees of freedom. We construct the
tight-binding model describing the glided helical edge by stacking the QSH
insulators. For definiteness we choose the Kane-Mele model\cite{KM} to
describe the QSH insulator in each layer, 
\begin{equation}
\hat{H}_{\text{QSH}}=-t\sum_{\left\langle i,j\right\rangle
s}c_{is,t}^{\dagger }c_{js,t}+i\frac{\lambda _{\text{SO}}}{3\sqrt{3}}%
\sum_{\left\langle \!\left\langle i,j\right\rangle \!\right\rangle s}s\nu
_{ij}c_{is,t}^{\dagger }c_{js,t},
\end{equation}%
where $s=\pm $ is the spin index. We propose the glided QSH insulator model
by 
\begin{align}
H_{\text{gQSH}}\left( \mathbf{k}\right) & =\Gamma _{\text{QSH}}\left(
k_{z}\right) +H_{\text{QSH}}\left( k_{x},k_{y}\right)  \notag \\
& +G_{x}\left( k_{z}\right) TH_{\text{QSH}}\left( k_{x},k_{y}\right)
T^{-1}G_{x}\left( k_{z}\right) ^{-1}.  \label{HamilGKM}
\end{align}%
The term $\Gamma _{\text{QSH}}\left( k_{x},k_{y}\right) $ represents the
interlayer coupling preserving the nonsymmorphic symmetry and the TRS.

The TRS is described by the operator $T=i\sigma _{y}K$ with $K$ the complex
conjugation. It is an antiunitary operator. The operation $G_{x}\left(
k_{z}\right) $\ is the successive operation of the half translation $t\left(
k_{z}\right) $\ and the mirror operation $M_{x}$. The mirror operation is
the composite operation of the reflection $\mathcal{R}_{x}$ and the spin
reversion in the presence of the spin. Thus, 
\begin{equation}
G_{x}\left( k_{z}\right) =i\sigma _{x}\otimes e^{-i\frac{ck_{z}}{2}}\Omega
(k_{z})\mathcal{R}_{x},
\end{equation}%
where $\sigma _{\alpha }$ is the Pauli matrix for the spin, and $\Omega
(k_{z})$ given by Eq.(\ref{Omega}) involves the pseudospin. It follows that $%
G_{x}^{2}\left( k_{z}\right) =-e^{-ick_{z}}\sigma _{0}\otimes \eta _{0}$.
The sign is negative due to the $2\pi $ spin rotation, which is different
from the spinless case.

We determine $\Gamma _{\text{QSH}}\left( k_{z}\right) $. We consider the
simplest candidates $t_{+}\cos \frac{ck_{z}}{2}\sigma _{\alpha }\otimes
\Omega (k_{z})$ and $t_{-}\sin \frac{ck_{z}}{2}\sigma _{\alpha }\otimes
\Omega (k_{z})$ describing the ordinary and skew interlayer hoppings as in
the case of the glided QAH insulator model. We choose the relevant terms by
requiring the symmetry property. Since the Kane-Mele model has the TRS, we
require the TRS also for $\Gamma _{\text{QSH}}\left( k_{z}\right) $. Namely,
we require both $T\Gamma _{\text{QSH}}\left( k_{z}\right) T^{-1}=\Gamma _{%
\text{QSH}}\left( -k_{z}\right) $ and $G_{x}\left( k_{z}\right) \Gamma _{%
\text{QSH}}\left( k_{z}\right) G_{x}^{-1}\left( k_{z}\right) =\Gamma _{\text{%
QSH}}\left( k_{z}\right) $. The only possible term is 
\begin{equation}
\Gamma _{\text{QSH}}\left( k_{z}\right) =\left( t_{+}\sigma _{0}\cos \frac{%
ck_{z}}{2}+t_{-}\sigma _{x}\sin \frac{ck_{z}}{2}\right) \otimes \Omega
\left( k_{z}\right) .  \label{GammaKM}
\end{equation}
We obtain 
\begin{equation}
G_{x}\left( k_{z}\right) H_{\text{gQSH}}\left( k_{x},k_{y},k_{z}\right)
G_{x}^{-1}\left( k_{z}\right) =H_{\text{gQSH}}\left(
-k_{x},k_{y},k_{z}\right) .
\end{equation}%
We focus on the surface made of the edges of each layers, where $k_{y}=0$
and we set $k=k_{x}$.

\begin{figure}[t]
\centerline{\includegraphics[width=0.49\textwidth]{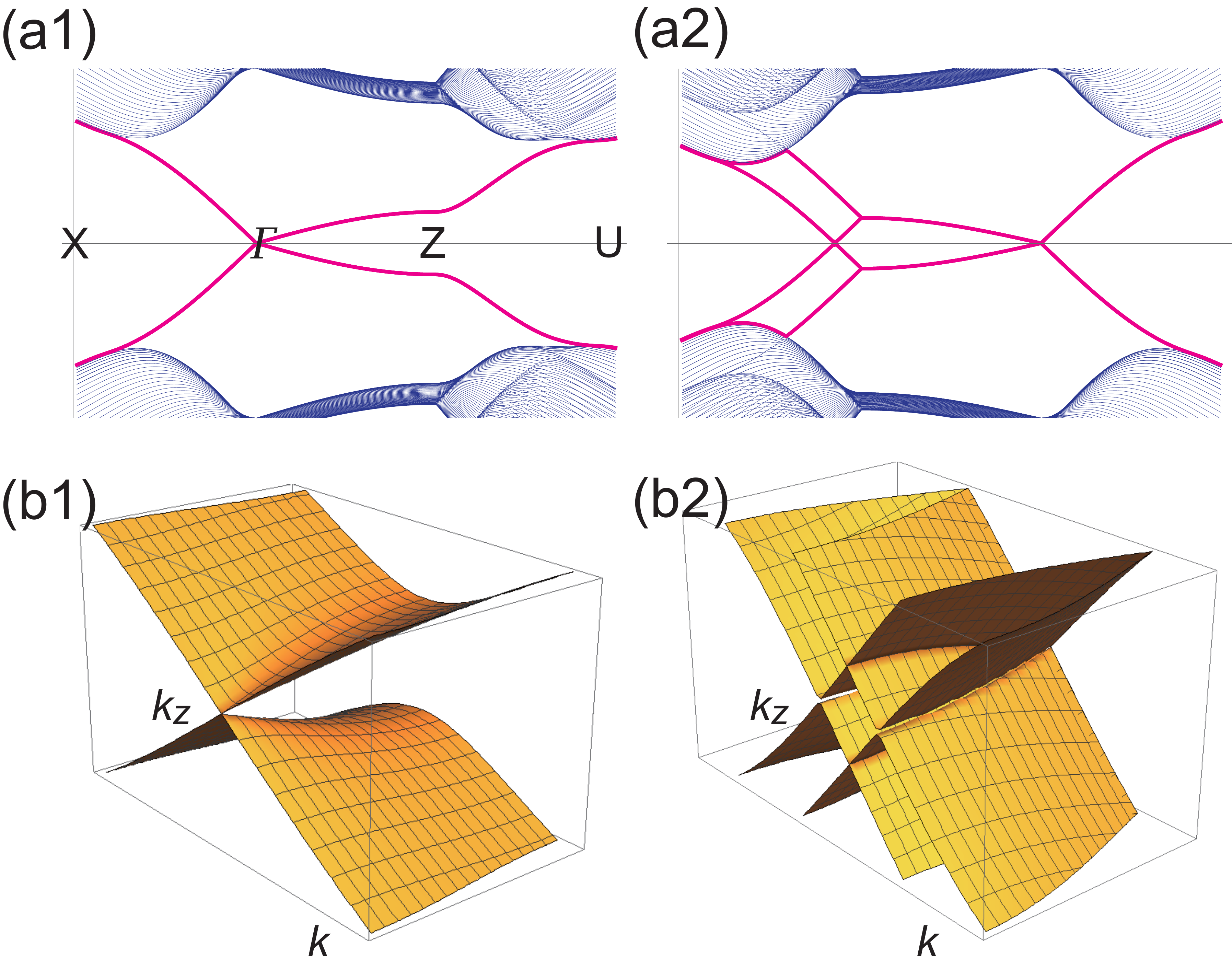}}
\caption{Band structure of the glided QSH insulator model for $t_{+}t_{-}=0$%
. The band structure along the X$\Gamma $ZU line for (a1) $t_{+}=0$, and
(a2) $t_{-}=0$. The red curves near the Fermi level represent the surface
states. (b1),(b2) Bird's eye's views of the energy spectrum for the surface
states.}
\label{FigGlass}
\end{figure}

We plot the band structure based on the tight-binding model Eq.(\ref%
{HamilGKM}) together with Eq.(\ref{GammaKM}) for $t_{+}t_{-}\neq 0$ in Fig.%
\ref{FigMain} and for $t_{+}t_{-}=0$ in Fig.\ref{FigGlass}. First of all,
the energy dispersion for the surface mode [Fig.\ref{FigMain}(a)] reproduces
excellently the results obtained by the DFT theory\cite{Hour} and the ARPES\
experiment\cite{KHg}. It is remarkable that the hourglass state emerges due
to this interlayer hopping [Fig.\ref{FigMain}(b)]. It should be noted that
the band crossing occurs irrespective to the sign of $t_{+}t_{-}$ since
these four bands connect between $k_{z}=0$ and $k_{z}=\pi /c$ in contrast to
the spinless case. There is the Kramers degeneracy at the time-reversal
invariant momentum $\Gamma $ ($k=0$ and $k_{z}=0$). On the other hand, there
is a double-fold degeneracy for all $k$ with $k_{z}=\pi /c$. This is the
pseudo-Kramers degeneracy of the operator\cite{Hour,Coho} $G_{x}\left(
k_{z}\right) T$ at $k_{z}=\pi /c$ due to the fact $\left( G_{x}\left( \pi
/c\right) T\right) ^{2}=-1$. Namely, the anti-unitary operator $G_{x}\left(
\pi /c\right) T$ acts like the time-reversal symmetry and assures the
pseudo-Kramers doublet.

When $t_{+}=0$, the two Kramers degeneracies at $E=\pm t_{+}$ with $k_{z}=0$
become identical and becomes a four-fold degenerated state, whose band
structure is plotted in Fig.\ref{FigGlass}(a1) and (b1). On the other hand,
when $t_{-}=0$, the two pseudo-Kramers degeneracies at $E=\pm t_{-}$ with $%
k_{z}=\pi /c$ becomes identical and becomes to a four-fold degenerated
state, whose band structure is plotted in Fig.\ref{FigGlass}(a2) and (b2).\ 

\section{Glided Dirac theory of helical edges}

We can prove these properties of the band structure depending on $t_{+}$\
and $t_{-}$\ analytically based on the Dirac theory. The Dirac theory is
derived to describe the surface state, 
\begin{equation}
H_{\text{gQSH}}\left( k,k_{z}\right) =vk\sigma _{z}+\Gamma _{\text{QSH}%
}\left( k_{z}\right)
\end{equation}%
together with Eq.(\ref{GammaKM}). The energy spectrum is given by%
\begin{equation}
E\left( k,k_{z}\right) =\pm \sqrt{\left( \sqrt{v^{2}k^{2}+\left( t_{+}\sin 
\frac{ck_{z}}{2}\right) ^{2}}\pm t_{-}\cos \frac{ck_{z}}{2}\right) ^{2}}.
\end{equation}%
We plot the energy spectrum along the X$\Gamma $ZU line in Fig.\ref{FigMain}
(c). This analytical result reproduces remarkably well the spectrum obtained
based on the tight-binding model [Fig.\ref{FigMain}(a)], the DFT theory\cite%
{Hour} and the ARPES\ experiment\cite{KHg}. Typical featues read as follows:

(i) The energy spectrum along the $\Gamma $Z line ($k=0$) is given by $%
E\left( 0,k_{z}\right) $: The gap closes at $\tan \frac{ck_{z}}{2}=\pm
t_{0}/t_{x}$. This band crossing is protected by the nonsymmorphic symmetry
and the chiral symmetry as in the case of the spinless model, where the
glide eigenvalue is given by $g_{\pm }\left( k_{z}\right) =\pm
ie^{-ick_{z}/2}$ and the chiral operator is given by $C=\sigma _{x}\eta _{z}$%
.

(ii) Along the ZU line ($k_{z}=\pi /c$), the energy spectrum is given by $%
E_{\pm }\left( k,\pi /c\right) $ with the double-fold degeneracy for all $k$%
, which is the pseudo-Kramers doublet as discussed in the tight-binding
theory. The glide eigenvalues are $g_{x}\left( \pi /c\right) =\pm 1$, as
plotted in Fig.\ref{FigMain}(c).

(iii) Along the $\Gamma $X line ($k_{z}=0$), the energy spectrum is given by
the four linear edge states; $E_{\pm }\left( k,0\right) =vk\pm t_{0},-vk\pm
t_{0}$. Each edge states are index by the glide eigenvalue of $g_{x}\left(
0\right) =\pm i$, as plotted in Fig.\ref{FigMain}(c). There are the Kramers
degeneracy at the time-reversal invariant momentum $\Gamma $ ($k=0$ and $%
k_{z}=0$) with $E_{\pm }\left( 0,0\right) =\pm t_{0}$. On the other hand,
the energy splits for $k\neq 0$.

\begin{figure*}[t]
\centerline{\includegraphics[width=0.69\textwidth]{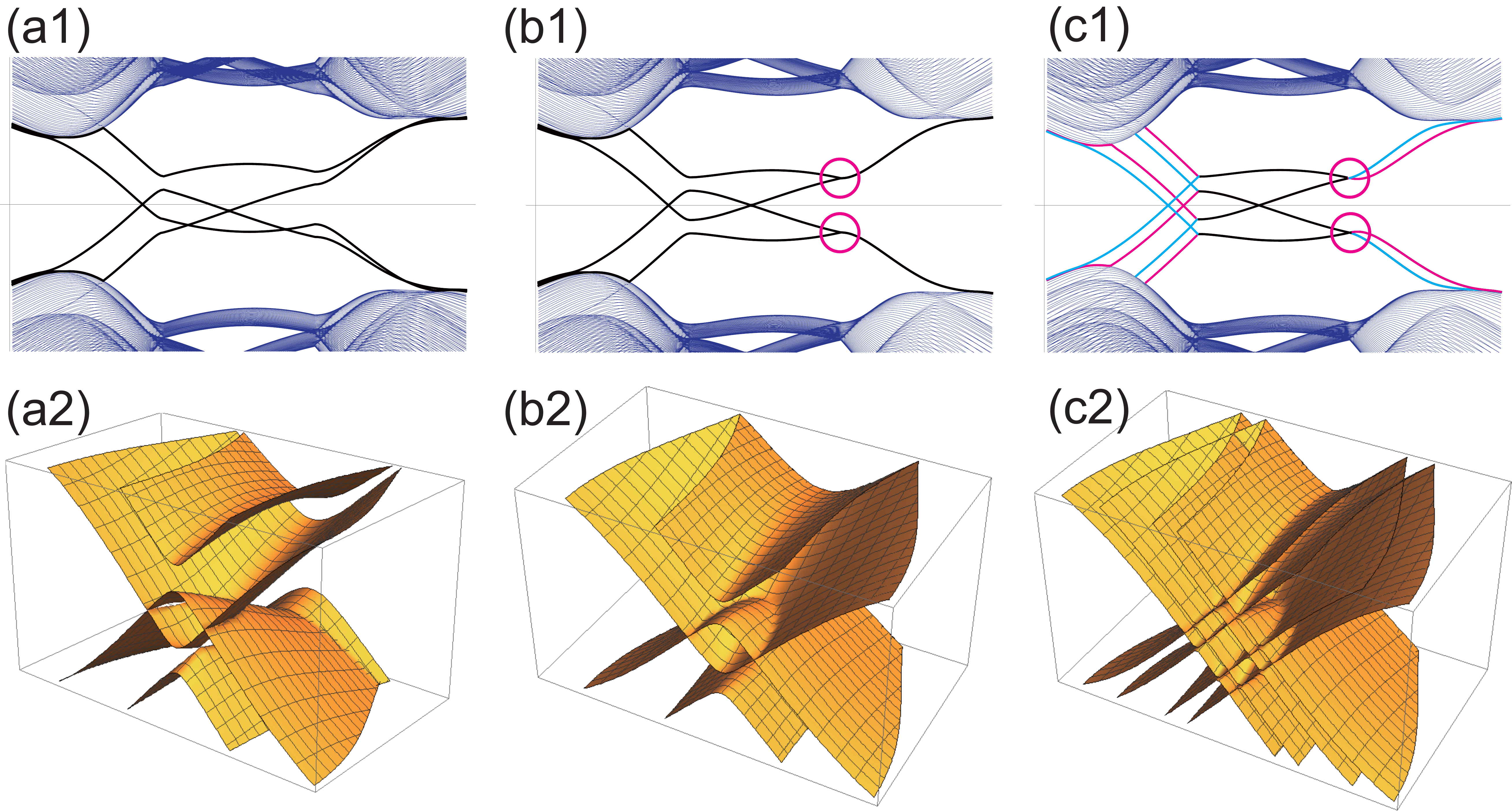}}
\caption{Hourglass fermion surface state along the X$\Gamma $ZU axis without
the TRS. The TRS is broken by applying the magnetic field. (a) $B_{x}\neq 0$%
, (b) $B_{y}\neq 0$ and (c) $B_{z}\neq 0$. (a1)$\sim $(c1) The surface band
structures become different between the both side of the edges. That of the
one side is colored in magenta, while that of the other side is colored in
cyan. (a2)$\sim $(c2) Bird's eye's views of the energy spectrum for the
surface states.}
\label{FigBz}
\end{figure*}

\section{Breaking the hourglass state}

We apply external magnetic field to the sample by introducing the Zeeman
coupling $\mathbf{B}\cdot \mathbf{\sigma }$. The band structures are
illustrated in Fig.\ref{FigBz}, which are interpreted based on the symmetry
as follows. When the magnetic field is along $y$ or $z$ direction, the $%
G_{x}T $ symmetry is preserved although the nonsymmorphic symmetry $G_x$ and
the TRS are broken. As a result, while the pseudo-Kramers degeneracy at $%
k_{z}=\pi /c$ is preserved, the Kramers degeneracy at $k_{z}=0$ is broken.
Although the band crossing is preserved even after the introduction of the
additional terms $\sigma_y$ or $\sigma_z$, it is not protected by the
nonsymmorphic symmetry $G_x$. The connection of the edge states along $%
k_{z}=0$ line is different between $B_{y}\sigma _{y}$ and $B_{z}\sigma _{z}$%
, as shown in Fig.\ref{FigBz}(b) and (c). We note that the degeneracy
between the both sides of the edges is broken due to the $\sigma _{z}$ term,
where $8$ bands emerge. The Zeeman field is opposite between the two sides
of the surfaces, which results in the difference between the surface band
structure between the both sides of the surfaces. On the other hand, $%
B_{x}\sigma _{x}$ breaks both the $G_{x}T$ and $T$ symmetries, which results
in the breaking both of the Kramers and pseudo Kramers degeneracies, as
shown in Fig.\ref{FigBz}(a). The band structure is not symmetric with
respect to the Fermi energy. This is because that the chiral symmetry $%
C=\sigma _{x}\eta _{z}$ is also broken, which results in the shift of the
gap closing point away from the Fermi energy with $B_{x}\sigma _{x}$.
However the band crossing is protected by the nonsymmorphic symmetry $G_x$
since the nonsymmorphic symmetry is preserved.

\begin{figure*}[t]
\centerline{\includegraphics[width=0.8\textwidth]{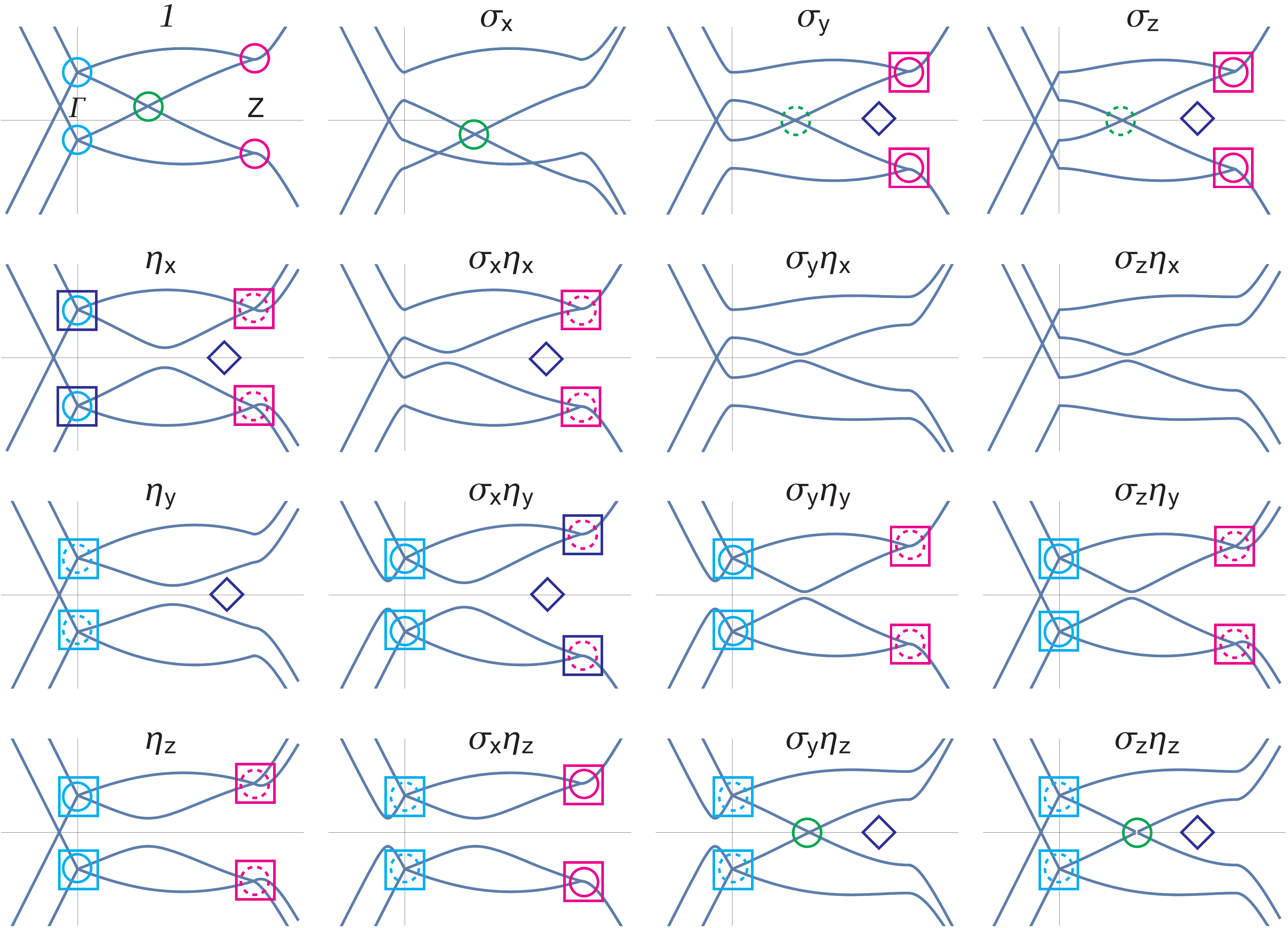}}
\caption{Energy spectrum of the hourglass fermion with the perturbation $V$
based on the Dirac theory. The band crossings protected by the glide
symmetry $G_{x}$ occur for $V\propto 1$, $\protect\sigma _{x}$, $\protect%
\sigma _{y}\protect\eta _{z}$, $\protect\sigma _{z}\protect\eta _{z}$ and
are marked by solid green circles, while those which are not protected by $%
G_{x}$ are marked by dotted green circles. The energy degeneracies protected
by the time-reversal symmetry $T$ occur at the $\Gamma $ point for $V\propto
1$, $\protect\eta _{x}$, $\protect\sigma _{x}\protect\eta _{y}$, $\protect%
\sigma _{y}\protect\eta _{y}$, $\protect\sigma _{z}\protect\eta _{y}$, $%
\protect\eta _{z}$ and are denoted by solid cyan circles, while those which
are not protected by $T$ are marked by the dotted cyan circles. The energy
degeneracies protected by the local chiral symmetry $C_{0}$ occurs for $%
V\propto \protect\eta _{y}$, $\protect\sigma _{x}\protect\eta _{y}$, $%
\protect\sigma _{y}\protect\eta _{y}$, $\protect\sigma _{z}\protect\eta _{y}$%
, $\protect\eta _{z}$, $\protect\sigma _{x}\protect\eta _{z}$, $\protect%
\sigma _{y}\protect\eta _{z}$, $\protect\sigma _{z}\protect\eta _{z}$ and
are marked by solid cyan squares. The energy degeneracies at the $Z$ point
protected by the time-reversal nonsymmorphic symmetry $G_{x}T$ occurs for $%
V\propto 1$, $\protect\sigma _{y}$, $\protect\sigma _{z}$, $\ \protect\sigma %
_{x}\protect\eta _{z}$ are marked by solid magenta circles, while those
which are not protected by $G_{x}T$ are marked by dotted magenta circles.
The energy degeneracies protected by the local chiral symmetry $C_{\protect%
\pi }$ occurs for $V\propto \protect\sigma _{y}$, $\protect\sigma _{z}$, $%
\protect\eta _{x}$, $\protect\sigma _{x}\protect\eta _{x}$, $\protect\sigma %
_{y}\protect\eta _{y}$, $\protect\sigma _{z}\protect\eta _{y}$, $\protect%
\eta _{z}$, $\protect\sigma _{x}\protect\eta _{z}$ and are marked by solid
magenta squares. The degeneracies occur due to the fact $V\propto H_{\text{%
gQSH}}$ at the $\Gamma $ point for $\protect\eta _{x}$ and at the $Z$ point
for $\protect\sigma _{x}\protect\eta _{y}$, and are marked by solid purple
squares. Symmetric band structures along $E=0$ due to the chiral symmetry
for $V\propto \protect\sigma _{y}$, $\protect\sigma _{x}\protect\eta _{x}$, $%
\protect\eta _{y}$, $\protect\sigma _{z}\protect\eta _{z}$ are marked by
solid purple diamonds.}
\label{HS}
\end{figure*}

We further investigate how the band structures are modified by introducing
perturbation terms of the form $V=V_{0}\sigma _{\alpha }\eta _{\beta }$. See
Section \ref{SecSymmetry} for details. We illustrate the results in Fig.\ref%
{HS}, which are interpreted based on the symmetry as follows. The degeneracy
at the $\Gamma $ and $Z$ points are well explained by the symmetry
operations $T $ and $G_{x}T$. However it is interesting that there are
degeneracies at the $\Gamma $ and $Z$ points which are not protected by
these symmetries. In order to clarify these degeneracies we write down the
Hamiltonian at the $\Gamma $ and $Z$ points, 
\begin{equation}
H_{\text{gQSH}}\left( 0,0\right) =t_{+}\sigma _{0}\eta _{x},\quad H_{\text{%
gQSH}}\left( 0,\pi /c\right) =t_{-}\sigma _{x}\eta _{y}.
\end{equation}%
When $\left[ H_{\text{gQSH}},V\right] \neq 0$, the energy spectrum of the
Hamiltonian $H_{\text{gQSH}}+V$ is given by the two two-fold degenerate
levels,%
\begin{equation}
E=\pm \sqrt{t_{\pm }^{2}+V_{0}^{2}}.
\end{equation}%
On the other hand, when $\left[ H_{\text{gQSH}},V\right] =0$, it is given by
non-degenerate four levels,%
\begin{equation}
E=\pm t_{\pm }\pm V_{0}.
\end{equation}%
Hence, the degeneracy is assured by $\left[ H_{\text{gQSH}},V\right] \neq 0$%
. It is identical to the condition $\left\{ H_{\text{gQSH}},V\right\} =0$,
which is a local chiral symmetry, in the present case of $V=V_{0}\sigma
_{\alpha }\eta _{\beta }$. The local chiral symmetry at the $\Gamma $ point
is explicitly written as%
\begin{equation}
C_{0}H\left( 0,0\right) +H\left( 0,0\right) C_{0}=0
\end{equation}%
with%
\begin{equation}
C_{0}=\eta _{x},
\end{equation}%
which protects the symmetry along $E=0$ at the $\Gamma $ point, as marked by
solid cyan squares in Fig.\ref{HS}. Similarly, the local chiral symmetry at
the $Z$ point is explicitly written as%
\begin{equation}
C_{\pi }H\left( 0,\pi /c\right) +H\left( 0,\pi /c\right) C_{\pi }=0
\end{equation}%
with%
\begin{equation}
C_{\pi }=\sigma _{x}\eta _{y},
\end{equation}%
which protects the symmetry along $E=0$ at the $Z$ point, as marked by solid
magenta squares in Fig.\ref{HS}. A exception occurs provided the
perturbation term $V$ is proportional to $H_{\text{gQSH}}$, where the energy
spectrum is given by the two two-fold degenerate levels,%
\begin{equation}
E=t_{\pm }+V_{0},\quad -t_{\pm }-V_{0},
\end{equation}%
although the original Hamiltonian $H_{\text{gQSH}}$ and $V$ commute
trivially, $\left[ H_{\text{gQSH}},V\right] =0$. We mark this case by solid
purple squares in Fig.\ref{HS}.

\section{Symmetry analysis}

\label{SecSymmetry} We investigate the symmetry of the additional
perturbation term $V$ and its effects on the band structure in detail, where 
$V$ is of the form%
\begin{equation}
V=\sigma _{\alpha }\eta _{\beta }.
\end{equation}%
The glide symmetry $G_{x}$ is given by 
\begin{equation}
G_{x}\left( k_{z}\right) =i\sigma _{x}\otimes e^{-i\frac{ck_{z}}{2}}\Omega
(k_{z})\mathcal{R}_{x},
\end{equation}%
and characterized by the action%
\begin{equation}
G_{x}\left( k_{z}\right) H\left( k,k_{z}\right) G_{x}^{-1}\left(
k_{z}\right) =H\left( -k,k_{z}\right)
\end{equation}%
on the unperturbed Hamiltonian $H\left( k,k_{z}\right) $. The perturbation
term $V$ is classified by the glide symmetry as 
\begin{equation}
G_{x}\left( k_{z}\right) V\left( -k,k_{z}\right) G_{x}^{-1}\left(
k_{z}\right) =\varepsilon ^{G_{x}}V\left( -k,k_{z}\right) ,
\end{equation}%
where $\varepsilon ^{G_{x}}=\pm 1$; we give $\varepsilon ^{G_{x}}$ for
various $V$ in Table I. The glide symmetry is preserved (violated) for $%
\varepsilon ^{G_{x}}=1$ ($-1$). The glide symmetry protects the gap closing
for the perturbations $1$, $\sigma _{x}$, $\sigma _{y}\eta _{z}$, $\sigma
_{z}\eta _{z}$, as marked by solid green circles in Fig.\ref{HS}.

The TRS $T$ is defined by 
\begin{equation}
TH\left( k,k_{z}\right) T^{-1}=H\left( -k,-k_{z}\right)
\end{equation}%
with 
\begin{equation}
T=i\sigma _{y}K.
\end{equation}%
The TRS protects the degeneracies at $k=0$ and $k_{z}=0$ for the
perturbations $1$, $\eta _{x}$, $\sigma _{x}\eta _{y}$, $\sigma _{y}\eta
_{y} $, $\sigma _{z}\eta _{y}$, $\eta _{z}$, as marked by solid cyan circles
in Fig.\ref{HS}. The perturbation term $V$ is classified by the TRS as 
\begin{equation}
TV\left( k,k_{z}\right) T^{-1}=\varepsilon ^{T}V\left( -k,-k_{z}\right) ,
\end{equation}%
where $\varepsilon ^{T}=\pm 1$; we give $\varepsilon ^{T}$ for various $V$
in Table I.

\begin{table*}[t]
\begin{tabular}{|c|c|c|c|c|c|c|c|c|c|c|c|c|c|c|c|c|}
\hline
& $1$ & $\sigma _{x}$ & $\sigma _{y}$ & $\sigma _{z}$ & $\eta _{x}$ & $%
\sigma _{x}\eta _{x}$ & $\sigma _{y}\eta _{x}$ & $\sigma _{z}\eta _{x}$ & $%
\eta _{y}$ & $\sigma _{x}\eta _{y}$ & $\sigma _{y}\eta _{y}$ & $\sigma
_{z}\eta _{y}$ & $\eta _{z}$ & $\sigma _{x}\eta _{z}$ & $\sigma _{y}\eta
_{z} $ & $\sigma _{z}\eta _{z}$ \\ \hline
$T$ & $+$ & $-$ & $-$ & $-$ & $+$ & $-$ & $-$ & $-$ & $-$ & $+$ & $+$ & $+$
& $+$ & $-$ & $-$ & $-$ \\ \hline
$G_{x}$ & $+$ & $+$ & $-$ & $-$ & X & X & X & X & X & X & X & X & $-$ & $-$
& $+$ & $+$ \\ \hline
$G_{x}T$ & $+$ & $-$ & $+$ & $+$ & X & X & X & X & X & X & X & X & $-$ & $+$
& $-$ & $-$ \\ \hline
$G_{y}$ & $+$ & $-$ & $+$ & $-$ & X & X & X & X & X & X & X & X & $-$ & $+$
& $-$ & $+$ \\ \hline
$C$ & $-$ & $-$ & $+$ & $+$ & $+$ & $+$ & $-$ & $-$ & $+$ & $+$ & $-$ & $-$
& $-$ & $-$ & $+$ & $+$ \\ \hline
$C_{0}$ & $-$ & $-$ & $-$ & $-$ & $-$ & $-$ & $-$ & $-$ & $+$ & $+$ & $+$ & $%
+$ & $+$ & $+$ & $+$ & $+$ \\ \hline
$C_{\pi }$ & $-$ & $-$ & $+$ & $+$ & $+$ & $+$ & $-$ & $-$ & $-$ & $-$ & $+$
& $+$ & $+$ & $+$ & $-$ & $-$ \\ \hline
\end{tabular}%
\caption{{The sign of the symmetry operations $\protect\varepsilon ^{S}=\pm $
with $S=T,G_{x},G_{x}T,C,C_{0},C_{\protect\pi }$. X denotes that there is no
symmetry. $\protect\varepsilon ^{S}=+$ (}$-$) {indicates that the symmetry }$%
S$ is preserved (violated) by the perturbation,.}
\label{TAB}
\end{table*}

The time-reversal glide symmetry $G_{x}\left( k_{z}\right) T$\ is defined by
the product of the glide and time-reversal symmetries as%
\begin{equation}
\left( G_{x}\left( k_{z}\right) T\right) H\left( k,k_{z}\right) \left(
G_{x}\left( k_{z}\right) T\right) ^{-1}=H\left( k,-k_{z}\right) .
\end{equation}%
The signs of the combined operation $\varepsilon ^{G_{x}T}$ are given by the
product of the glide and the time-reversal symmetry 
\begin{equation}
\varepsilon ^{G_{x}T}=\varepsilon ^{G_{x}}\varepsilon ^{T}.
\end{equation}%
This symmetry is anti-unitary and leads to the pseudo-Kramers degeneracy at $%
Z$ point for the perturbations $1$, $\sigma _{y}$, $\sigma _{z}$, $\ \sigma
_{x}\eta _{z}$, as marked by solid magenta circles in Fig.\ref{HS}.

The particle-hole symmetry $P$ is defined by%
\begin{equation}
P=i\sigma _{z}\eta _{z}K
\end{equation}%
with%
\begin{equation}
PH\left( k,k_{z}\right) P^{-1}=-H\left( -k,k_{z}\right) .
\end{equation}%
The chiral symmetry $C$\ is the product of the TRS and the PHS, and defined
by%
\begin{equation}
CH\left( k,k_{z}\right) +H\left( k,k_{z}\right) C=0
\end{equation}%
with%
\begin{equation}
C=\sigma _{x}\eta _{z}.
\end{equation}%
The perturbation term $V$ is classified with by the chiral symmetry as 
\begin{equation*}
CV\left( k,k_{z}\right) +\varepsilon ^{C}V\left( k,k_{z}\right) C=0,
\end{equation*}%
where $\varepsilon ^{C}=\pm 1$; we give $\varepsilon ^{C}$ for various $V$
in Table I. The band is symmetric along $E=0$ if there is the chiral
symmetry for the perturbations $\sigma _{y}$, $\sigma _{x}\eta _{x}$, $\eta
_{y}$, $\sigma _{z}\eta _{z}$, as marked by solid purple diamonds in Fig.\ref%
{HS}. However it seems that there are many band structures which are
symmetric along $E=0$, as marked in the dashed purple diamonds. In order to
clarify this symmetry, we further define the local chiral symmetry $C_{0}$
at the $\Gamma $ point, 
\begin{equation}
C_{0}H\left( 0,0\right) +H\left( 0,0\right) C_{0}=0
\end{equation}%
with%
\begin{equation}
C_{0}=\eta _{x},
\end{equation}%
which protects the symmetry along $E=0$ at the $\Gamma $ point for the
perturbations $\eta _{y}$, $\sigma _{x}\eta _{y}$, $\sigma _{y}\eta _{y}$, $%
\sigma _{z}\eta _{y}$, $\eta _{z}$, $\sigma _{x}\eta _{z}$, $\sigma _{y}\eta
_{z}$, $\sigma _{z}\eta _{z}$, as marked by solid cyan squares in Fig.\ref%
{HS}. We note that $C_{0}$ is proportional to $H(0,0)$. The perturbation
term $V$ is classified by the chiral symmetry as 
\begin{equation}
C_{0}V+\varepsilon ^{C_{0}}VC_{0}=0,
\end{equation}%
where $\varepsilon ^{C_{0}}=\pm 1$; we give $\varepsilon ^{C_{0}}$ for
various $V$ in Table I. Similarly, we define the local chiral symmetry $%
C_{\pi }$ at the $Z$ point, 
\begin{equation}
C_{\pi }H\left( 0,\pi /c\right) +H\left( 0,\pi /c\right) C_{\pi }=0
\end{equation}%
with%
\begin{equation}
C_{\pi }=\sigma _{x}\eta _{y},
\end{equation}%
which protects the symmetry along $E=0$ at the $Z$ point for the
perturbations $\sigma _{y}$, $\sigma _{z}$, $\eta _{x}$, $\sigma _{x}\eta
_{x}$, $\sigma _{y}\eta _{y}$, $\sigma _{z}\eta _{y}$, $\eta _{z}$, $\sigma
_{x}\eta _{z}$, as marked by solid magenta squares in Fig.\ref{HS}. We note
that $C_{0}$ is proportional to $H(0,\pi /c)$. The perturbation term $V$ is
classified by the chiral symmetry as 
\begin{equation}
C_{\pi }V+\varepsilon ^{C_{\pi }}VC_{\pi }=0
\end{equation}%
where $\varepsilon ^{C_{\pi }}=\pm 1$; we given $\varepsilon ^{C_{\pi }}$
for various $V$ in Table I.

\section{Discussion}

We have demonstrated that the hourglass fermion surface state is well
described by the glided QSH insulator model and the glided Dirac theory,
where the degeneracy at the high-symmetry points are protected by the
nonsymmorphic symmetry and the TRS. The symmetry with respect to the Fermi
energy is protected by the chiral symmetry. We have shown that these
degeneracies are lifted by applying magnetic field, which will be observable
in ARPES experiments. Although the hourglass fermion surface state was found
in a specific material, KHgSb, both theoretically and experimentally, our
results show that it is universal in the helical edge system with the
nonsymmorphic symmetry.

\acknowledgements

The author is very much grateful to N. Nagaosa for many helpful discussions
on the subject. He thanks the support by the Grants-in-Aid for Scientific
Research from MEXT KAKENHI (Grant Nos.JP25400317 and JP15H05854).


\begin{thebibliography}{99}
\bibitem{Hasan} M. Z. Hasan and C. L. Kane, Rev. Mod. Phys. 82, 3045 (2010)

\bibitem{Qi} X.-L. Qi and S.-C. Zhang, Rev. Mod. Phys. 83, 1057 (2011)

\bibitem{Haldane} F. D. M. Haldane, Phys. Rev. Lett. 61, 2015 (1988)

\bibitem{Chang} C.-Z. Chang. et.al., Sicence 340, 167 (2013)

\bibitem{KM} C. L. Kane and E. J. Mele, Phys. Rev. Lett. 95, 146802 (2005):
C. L. Kane and E. J. Mele, Phys. Rev. Lett. \textbf{95}, 226801 (2005).

\bibitem{B1} B. A. Bernevig, T. L. Hughes, S.-C. Zhang, Science, 314, 1757
(2006)

\bibitem{B2} B. A. Bernevig and S.-C. Zhang, Phys. Rev. Lett. 96, 106802
(2006)

\bibitem{TCI} L. Fu, Phys. Rev. Lett. 106, 106802 (2011).

\bibitem{Hsieh} T. H. Hsieh, et.al, Nat. Commun. 3, 982 (2012).

\bibitem{Tanaka} Y. Tanaka, et.al., Nat. Phys. 8, 800 (2012).

\bibitem{Dziawa} P. Dziawa, et.al., Nat. Mater. 11, 1023 (2012).

\bibitem{Xu} S.-Y. Xu, et al., Nat. Commun. 3, 1192 (2012).

\bibitem{Robert} R.-J. Slager, A. Mesaros, V. Juricic, J. Zaanen, Nature
Physics 9, 98?102 (2013)

\bibitem{Ando} Y. Ando and L. Fu, Annual Review of Condensed Matter Physics,
6, 361 (2015).

\bibitem{Aris} A. Alexandradinata, C. Fang, M. J. Gilbert, and B. A.
Bernevig, Phys. Rev. Lett. 113, 116403 (2014)

\bibitem{Van} C.-X. Liu, R.-X. Zhang, and B. K. VanLeeuwen, Phys. Rev. B 90,
085304 (2014).

\bibitem{Para} S. A. Parameswaran, A. M. Turner, D. P. Arovos, and A.
Vishwanath, Nat. Phys. 9, 299 (2013).

\bibitem{Fang} C. Fang and L. Fu, Phys. Rev. B 91, 161105 (2015).

\bibitem{Dong} X.-Y. Dong and C.-X. Liu, Phys. Rev. B 93, 045429 (2016)

\bibitem{Young} S. M. Young and C. L. Kane, Phys. Rev. Lett. 115, 126803
(2015)

\bibitem{Watanabe} H. Watanabe, H. C. Po, A. Vishwanath, M. P. Zaletel,
Proc. Natl. Acad. Sci. 112, 14551 (2015)

\bibitem{Po} H. C. Po, H. Watanabe, M. P. Zaletel, A. Vishwanath, Sci. Adv.
2(4), e1501782 (2016)

\bibitem{BJYang} B.-J. Yang, T. A. Bojesen, T. Morimoto, A. Furusaki,
cond-mat/arXiv:1604.00843

\bibitem{Murakami} H. Kim and S. Murakami, Phys. Rev. B 93, 195138 (2016)

\bibitem{Sch} Y. Z. Zhao, A. P. Schnyder, cond-mat/arXiv:1606.03698

\bibitem{ShiozakiB} K. Shiozaki, M. Sato and K. Gomi, Phys. Rev. B 91,
155120 (2015)

\bibitem{Shiozaki} K. Shiozaki, M. Sato and K. Gomi, Phys. Rev. B 93, 195413
(2016)

\bibitem{Chen} Y. Chen, H.-S. Kim and H.-Y. Kee, Phys. Rev. B 93, 155140
(2016)

\bibitem{Varjas} D. Varjas, F. de Juan and Y.-M. Lu, Phys. Rev. B 92, 195116
(2015)

\bibitem{ChangErten} P.-Y. Chang, O. Erten, P. Coleman, cond-maat/arXiv:1603.03435

\bibitem{Hour} Z. Wang, A. Alexandradinata, R. J. Cava and B. A. Bernevig.
Nature 532, 189 (2016)

\bibitem{Coho} A. Alexandradinata, Z. Wang and B. A. Bernevig, Phys. Rev. X
6, 021008 (2016)

\bibitem{KHg} J.-Z. Ma, et. al., cond-mat/arXiv:1605.06824
\end{thebibliography}
\end{document}